\documentclass{article}
\usepackage[T1]{fontenc}
\usepackage[utf8]{inputenc}
\usepackage[cmu15798]{ismir} 
\usepackage{amsmath,cite,url}
\usepackage{graphicx}
\usepackage{color}

\title{DARC: Drum accompaniment generation with fine-grained rhythm control}

\oneauthor
  {Trey Brosnan}
  {Carnegie Mellon University\\\texttt{rbrosnan@andrew.cmu.edu}}

\def\authorname{R. Brosnan}

\begin{document}

\maketitle

\begin{abstract}
In music creation, rapid prototyping is essential for exploring and refining ideas, yet existing generative tools often fall short when users require both structural control and stylistic flexibility. Prior approaches in stem-to-stem generation can condition on other musical stems but offer limited control over rhythm, and timbre-transfer methods allow users to specify specific rhythms, but cannot condition on musical context. We introduce DARC, a generative drum accompaniment model that conditions both on musical context from other stems and explicit rhythm prompts such as beatboxing or tapping tracks. Using parameter-efficient fine-tuning, we augment STAGE \cite{stage}, a state-of-the-art drum stem generator, with fine-grained rhythm control while maintaining musical context awareness.
\end{abstract}

\section{Introduction}
In recent years, numerous works \cite{musiccongen, stemgen, stage, coco-mulla, jukedrummer, musicgenstem, singsong} have achieved high-quality, musically coherent accompaniment generation. However, these methods often lack fine-grained control over time-varying features. Such control is often desirable in the context of musical prototyping, where a creator wishes to quickly evaluate an early musical idea before investing substantial time into it. In this work, we focus on the Tap2Drum task, in which a user can record a rhythm prompt, such as a beatboxing or tapping track, and a generative model renders it as drums. State-of-the-art approaches for Tap2Drum focus on timbre transfer, where the user provides a timbre prompt to explicitly specify the desired drum timbre. For instance, \cite{tria} requires the user to provide drum audio as the timbre prompt; this can limit the speed of iteration, as different songs will require different drumkit sounds, and the user must search for an existing audio sample matching their desired timbre. Other works in music editing \cite{melodyFlow} provide text control, but it can be difficult to articulate drum timbres using text, and moreover these methods tend to suffer from timbre leakage \cite{tria}. Some works, both in Tap2Drum \cite{groovae, clavenet} and in accompaniment generation \cite{MusicControlNet,stage}, offer onset-based rhythm control, but this is too coarse to capture the implied timbre categories of a rhythm prompt.

We propose DARC, a drum accompaniment generation model that takes as input musical context and a rhythm prompt. Our rhythm feature representation, based on nonnegative matrix factorization (NMF), provides greater granularity than onset-based methods by classifying each onset into a timbre class. DARC is a fine-tuning of STAGE \cite{stage}, a SOTA drum accompaniment model. Our motivation for inferring timbre from musical context rather than a timbre prompt is twofold: first, drums are rarely a solo instrument, i.e. the end goal for a drum track is often to accompany a mix; second, removing the requirement for users to provide a timbre prompt can shorten their iteration cycle, enabling them to explore more ideas. For our dataset, we extract drum stems from the FMA dataset \cite{fma} using Demucs \cite{demucs1,demucs2}. During fine-tuning, we utilize the parameter-efficient method proposed in \cite{musiccongen}.

Our contributions are 2-fold:
\begin{itemize}
    \item We introduce a generative drum model that can condition on both musical context and specific rhythms, with timbre classes
    \item We evaluate our model on musical coherence with the input mix and onset and timbre class adherence to the rhythm prompt, exposing limitations in existing evaluation metrics
\end{itemize}

\begin{figure*}
    \centering
    \includegraphics[width=1\linewidth]{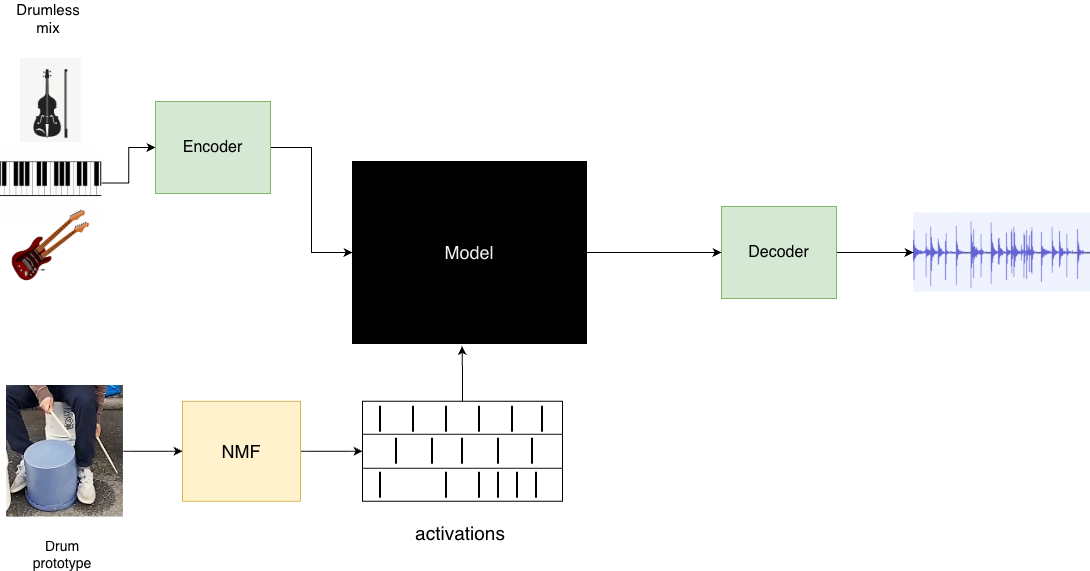}
    \caption{Architecture of the proposed rhythm-conditioned music generation model. Musical context and rhythm prompt are provided as audio inputs. The tokenized musical context is prepended to the input sequence, and the rhythm prompt is transcribed into (onset time, timbre class) pairs using non-negative matrix factorization (NMF). The rhythm embedding is passed through the self-attention layers via jump fine-tuning and adaptive in-attention \cite{musiccongen}. The model outputs EnCodec audio tokens that are decoded to the final waveform.}
    \label{fig:1}
\end{figure*}

\section{Related Work}\label{sec:related*works}

\subsection{Accompaniment Generation}
A recent line of work has explored music accompaniment generation \cite{subtractiveTraining,musicgenstem,stemgen,stage,musiccongen,singsong,jukedrummer}, which can generate one or more tracks to accompany given musical mix. Note that many of these models support text conditioning, and are in fact fine-tunings of the text-to-music model MusicGen \cite{musicGen}. While these stem-to-stem generation models can condition on other stems in the mix, they are not designed for fine-grained rhythm control. Some approaches allow for conditioning on onsets \cite{MusicControlNet,stage}. However, the rhythm control provided by these approaches is quite loose; the model does not preserve the onsets, but rather uses them as a guide to generate an embellished drum track. In addition, onset timings alone do not capture implied timbre classes, such as an onset being from a kick drum versus a snare. Our work seeks to provide tighter rhythm control and can preserve timbre classes.

Other work has focused on more specialized aspects of drum generation. For example, \cite{realtimeDrums} generates drum accompaniments in real time, and \cite{drumFills} uses a bidirectional language model to generate drum fills. We leave the adaptation of our methods for real-time or fill generation as future work.

\subsection{Tap2Drum Generation}
An alternative line of work explores the Tap2Drum task, which takes tapping or beatboxing as input and seeks to generate a drum track with the same rhythm. Tap2Drum was first introduced in \cite{groovae}, which takes onset times as input and generates drums as MIDI\footnote{This was released as an Ableton Live plugin: https://magenta.withgoogle.com/studio}. Other work such as TRIA \cite{tria} performs timbre transfer, directly converting the rhythm prompt audio to high-fidelity drum audio. In addition to a rhythm prompt, such methods take a timbre prompt in the form of audio, requiring users to present an audio sample with the exact timbre they desire. Further work has explored non-zero-shot timbre transfer \cite{demerle2024combining, demerle2024combiningaudiocontrolstyle, mancusi2025latentdiffusionbridgesunsupervised, engel2020ddsp, santos2023taps}, which requires re-training a model for each target timbre. Our model, DARC, generates a suitable timbre for the given input mix, avoiding the need to prompt or train for specific timbres. Moreover, our rhythm features encode timbre classes in addition to onsets times, providing greater granularity than existing timbre transfer approaches.

\section{Method}

\subsection{Overview}
Our model takes two audio-form inputs: a drumless mix as musical context and a rhythm prompt, such as a beatboxing or tapping track. Our goal is to generate a drum stem that faithfully maintains the onsets of each timbre class of the rhythm prompt while exhibiting strong musical coherence with the input mix. We fine-tune STAGE \cite{stage}, a recent open-source model that generates single stem accompaniments. STAGE itself is a fine-tuning of MusicGen \cite{musicGen}, using prefix-based conditioning on both drumless mixes and metronome-like pulse tracks during training. STAGE contains roughly 620M parameters; following \cite{musiccongen}, we use a parameter-efficient fine-tuning technique to reduce the trainable parameter count by an order of magnitude. Note that separate STAGE models were trained for drum and bass stems; we consider only the drum model in this work.
\subsection{Rhythm Feature Representation}\label{subsec:rhythm_features}
A key challenge in the Tap2Drum task is \textit{timbre leakage}: while the generated stem should exhibit close adherence to the rhythm prompt, its timbre should be independent of the rhythm prompt. To address this, we use non-negative matrix factorizaion (NMF) to obtain our rhythm features. NMF decomposes a magnitude spectrogram $S$ of a rhythm prompt into a product of matrices, $S=WH.$ The basis matrix $W$ encodes timbre information, and the activation matrix $H$ encodes timing information. In particular, the indices of the rows of $W$ and the columns of $H$ correspond to different timbre classes. To obtain our rhythm features, we ignore the matrix $W,$ leaving us with a matrix $H$ of the activation times of each timbre class. Hence, the rhythm-feature representation is MIDI-like: for a beatboxing track, it would contain the onset times and timbre-class indices of each note, but no information about the underlying vocal timbre. Crucially, we sort the timbre classes in decreasing order of total component energy, roughly corresponding to kick, snare, and hi-hat for the first three classes. This way, the model can identify the timbre classes without knowing the timbre information matrix $W.$

\subsection{Fine-Tuning}
Our base model, STAGE, is a MusicGen-Small model fine-tuned for generating drum stems conditioned on a drumless mix. During training, the authors prepended the input with the audio tokens of the drumless mix, followed by a delimiter token. Therefore, at inference time, the drum stem generation is framed as a continuation task, with the input mix as the prompt. The authors found this prefix-based conditioning method to be superior to cross-attention in their work \cite{stage}. We retain this mechanism for conditioning on the drumless mix, using a different approach to augment STAGE with fine-grained rhythm control.

During fine-tuning, we freeze approximately $80\%$ of the parameters of STAGE. First, we freeze the text encoder and audio token embedding modules. Then, we utilize two fine-tuning strategies proposed in \cite{musiccongen}: jump fine-tuning and adaptive in-attention. Under jump fine-tuning, only the first self-attention layer in each decoder block is fine-tuned, while the remaining three layers are frozen. In adaptive in-attention, the conditioning signal is reintroduced at the first layer of each block; this mechanism is applied to the first 75\% of the blocks. For example, for a decoder with 48 self-attention layers, we would have 12 self-attention blocks. All layers except $0,4,8,12,\ldots,44$ would be frozen, and the rhythm condition would be reapplied at layers $4,8,12,\ldots,32.$

For our dataset, we use FMA Small\cite{fma}, extracting drum stems using Demucs \cite{demucs1, demucs2}. We perform data augmentation on both the musical context and rhythm prompt, including tempo and pitch shifting, Guassian noise, and band-pass filtering, each applied independently with probability $0.25$. Any augmentation applied to the drumless mix is also applied the ground-truth drum stem during training to encourage consistency between the stem and the mix. We train on random 10-30 second chunks of audio, using log-uniform sampling to favor shorter lengths. This yields an average input length of $18.2$ seconds, corresponding to an expected duration of about 6 hours for the entire training set. Training was performed on an A100 GPU for 7 epochs with a batch size of 4, spanning 2 hours.

\section{Experimental Setup}
We compare our model against STAGE \cite{stage} and TRIA \cite{tria}, comparing audio quality, musical coherence, and rhythm prompt adherence, both overall and within particular timbre classes. We use the MUSDB18 dataset \cite{musdb18} for musical coherence and AVP Beatbox dataset \cite{avpBeatbox} for rhythm adherence.

\subsection{Audio Quality}\label{subsec:quality}
We personally evaluate audio quality in a subjective manner. Overall, we perceive the audio quality as quite poor, with frequent artifacts and non-drum instrument sounds in the background. We suspect that these issues originate from the stem separation step during our dataset creation. Errors in stem separation are known to manifest as bleed and artifacts \cite{bleed1}, which align with our observations. In future work, we wish to experiment with alternative stem separation models, as well as datasets that contain ground-truth stems, to evaluate this claim.

\subsection{Rhythm Prompt Adherence}\label{exp:rhythm}
We separate rhythm prompt adherence into timing accuracy, measured by Onset F1, and timbre class accuracy, measured by Kick and Snare F1. For onsets, we use a 70ms tolerance and perform onset detection on the generated and ground-truth stems using Beat-This \cite{beatThis}. For timbre class adherence, we use FrameRNN \cite{frameRNN} to transcribe the generated drum stems and compute the F1 score of the kick and snare onsets, using the standard 30ms and 100ms tolerances \cite{tria}, respectively. Note that while we attempted to transcribe the ground truth beatboxing tracks from AVP, the accuracy was extremely poor, and we instead used the ground-truth annotations provided by the dataset.

Due to audio quality issues discussed in \ref{subsec:quality} above, both the onset detection and drum transcription models demonstrated poor accuracy on DARC's outputs. Therefore, we post-processed our audio by gating the upper frequencies to reduce noise and bleed, enhancing transients, and applying light compression and normalization. For fair comparison, we applied the same post-processing to the ground truth rhythm prompts and all models being compared. Rhythm prompts were truncated to 9 seconds and rhythm prompts with less than 2 detected onsets were ignored (4 such files were found in AVP).

\subsection{Musical Coherence}
To evaluate musical coherence, we compute the COCOLA score \cite{cocola} between each drum stem and the drumless input mix. We use 10-second chunks of 50 random samples from MUSDB18 as our evaluation set. As a baseline, we compute the COCOLA score between the ground-truth drum stems and drumless mixes. To evaluate STAGE, we perform rhythm conditioning as described in the original paper \cite{stage}: we detect beats in the rhythm prompt and sum the corresponding click track with the musical context, using the result as the input to STAGE. For our model, we condition directly on the NMF rhythm features as described in \ref{subsec:rhythm_features}.

\section{Results and Discussion}
\subsection{Rhythm Prompt Adherence}
Table \ref{tab:rhythm_adherence} shows our rhythm adherence results. We observe that, across all three metrics, DARC is outperformed by TRIA and STAGE. As noted in \ref{subsec:quality} above, our model had very poor audio quality, which our evaluation models were not robust against. Even on the ground-truth rhythm prompts from the AVP dataset, these models displayed poor performance as discussed in Section \ref{exp:rhythm}. Furthermore, while our post-processing appeared qualitatively to improve the performance of the evaluation models, this was far from a perfect solution. In particular, we expect that if the audio quality of DARC were improved, with all else held constant, its experimental results would improve significantly. As such, improving the output audio fidelity is an important avenue for future work; we hypothesize that utilizing a GAN during training or altering our dataset, either by using a different source separator model or a dataset such as MoisesDB \cite{moisesdb} that contains ground-truth drum stems, could be effective methods.

\subsection{Musical Coherence}
Table \ref{tab:musical_coherence} shows our musical coherence results. We observe a significantly lower COCOLA score for DARC compared to STAGE and the ground truth. Again, we suspect that low audio fidelity (see Section \ref{subsec:quality}) may have played a role in these results. Interestingly, STAGE outperformed the ground truth in our experiment by a small margin. This is surprising, and while it's possible that STAGE simply generated more coherent drum stems than the ground truth, we believe that this instead reflects a limitation of the COCOLA model itself. Qualitatively, we observed that STAGE's outputs tended to be more embellished than the ground truth drum tracks, yielding a much greater number of total notes. We suspect that COCOLA rewarded STAGE for each note that was rhythmically coherent with the musical context, even when a human listener might view the embellishments as excessive. This provides motivation for future work to conduct human listening studies to evaluate musical coherence, as well as design musical coherence metrics that exhibit greater robustness to audio fidelity and alignment with human preferences.

\section{Conclusion}\label{sec:conclusion}
We proposed DARC, a drum accompaniment generation model that can be conditioned on a rhythm prompt in addition to musical context. Our NMF-based rhythm features allow for timbre class preservation without timbre leakage. While qualitatively, our model appeared to adhere to the rhythm prompt reasonably well, our quantitative results were underwhelming due to the poor audio fidelity of our outputs. This revealed a key limitation both of our model and of existing metrics for rhythm similarity and musical coherence. Future work could explore improving the audio quality of DARC; we propose either using alternative source separation models for dataset creation, or avoiding extraction completely by using datasets that contain ground-truth drum stems. Moreover, using a GAN during training might provide a mechanism to improve audio quality \cite{wavegan}. For the evaluation metrics, we encourage future work to explore robust rhythm adherence and musical coherence evaluation metrics that can handle various levels of audio fidelity. Finally, upon improvements to our model, we encourage future work to implement user-facing tools for DARC or other models that are designed to aid the music creation progress. By observing how human music creators interact with the technology, we can gain a more clear view of the real-world applicability of such models, as well as insights into broader impacts and areas for improvement.

\section{Broader Impacts}
Our model, DARC, is designed for co-creation with a human creator, allowing them to tightly control the rhythm profile of the generated output. However, we note that the timbre of the generated drum stem is decided by DARC based on the musical context, which represents a tradeoff for convenience versus control when compared to timbre transfer methods. At the same time, when compared to previous stem generation models such as STAGE, MusiConGen, StemGen, or MusicGen-Stem, we note that DARC accepts much more detailed rhythmic input; these works either take no rhythm conditioning, a BPM, or a click track as their rhythm input. Therefore, DARC lies somewhere between existing works for timbre transfer and stem generation in terms of user control.

In general, drum generation models have the potential to replace human drummers. Over time, they might result in fewer people learning to play physical drumsets, shifting musical culture away from human drummers. We note that DARC was designed for co-creation and rapid prototyping, but real-world usage can differ from initial intentions. As mentioned above, human interaction studies in future work can provide insights into real-world use-cases, and are a vital tool for analyzing broader impacts of DARC and other models.

Especially if models are not trained on sufficiently diverse datasets, they can exhibit bias toward certain musical styles or sounds, contributing to the homogenization of music. We note that our dataset, FMA Small, contains balanced levels of 8 different genres \cite{fma}, promoting diversity. However, most of the audio samples are Western music. Expanding DARC and other music AI works to non-Western music is an important avenue for future work, and can be challenging due to data scarcity.

\begin{table}[h]
  \centering
  \begin{tabular}{|l|c|c|c|}
    \hline
    Model & Onset F1 $\uparrow$ & Kick F1 $\uparrow$ & Snare F1 $\uparrow$ \\
    \hline
    STAGE & $0.270$ & $0.056$ & $0.134$\\
    TRIA & $0.347$ & $0.180$ & $0.382$\\
    \textbf{DARC} & $0.188$ & $0.053$ & $0.111$\\
    \hline
  \end{tabular}
  \caption{Rhythm prompt adherence results on the AVP \cite{avpBeatbox} dataset. Onset detection is performed by Beat-This \cite{beatThis}, and drum transcription is performed by FrameRNN \cite{frameRNN}. Onset, kick, and snare F1 scores are computed with tolerances of 70ms, 30ms and 100ms, respectively.}
  \label{tab:rhythm_adherence}
\end{table}

\begin{table}[h]
  \centering
  \begin{tabular}{|l|c|}
    \hline
    Model & COCOLA Score $\uparrow$ \\
    \hline
    STAGE & $63.9816$ \\
    \textbf{DARC} & $53.5908$\\
    \hline
    Ground-truth & $63.7227$ \\
    \hline
  \end{tabular}
  \caption{Musical coherence between the generated drum stem and input musical context, evaluated on 50 randomly selected tracks from MUSDB18 \cite{musdb18}.}
  \label{tab:musical_coherence}
\end{table}

\bibliography{ISMIRtemplate}

%
%
%
%

\end{document}